\documentclass[twocolumn]{revtex4}
\usepackage{amssymb,epsf}
\usepackage{latexsym}
\usepackage{xcolor}
\usepackage{epsfig}
\usepackage{float}
\usepackage{color}
\usepackage{amsmath}
\usepackage{bm}
\usepackage{amsthm}
\usepackage{amssymb}
\usepackage{amssymb,epsf}
\usepackage{latexsym}
\usepackage{epsfig}
\usepackage{graphicx}
\begin{document}
%%%%%%%%%%%%%%%%%%%%%%%%%%%%%%%%%%%%%%%%%%%%%%%%%%
\title{Generalized uncertainty principle and quantum non-locality}
%%%%%%%%%%%%%%%%%%%%%%%%%%%%%%%%%%%%%%%%%%%%%%%%%%%%%%%%%%%%%%%%%%
\author{$^1$S. Aghababaei\footnote{sarah.aghababaei@gmail.com}, $^2$H. Moradpour\footnote{hn.moradpour@maragheh.ac.ir}}
\address{$^1$ Department of Physics, Faculty of Sciences, Yasouj University, 75918-74934, Yasouj, Iran\\
$^2$ Research Institute for Astronomy and Astrophysics of Maragha
(RIAAM), University of Maragheh, P.O. Box 55136-553, Maragheh,
Iran}
%%%%%%%%%%%%%%%%%%%%%%%%%%%%%%%%%%%%%%%%%%%%%%%%%%%%%
\begin{abstract}
The emergence of the generalized uncertainty principle and the
existence of a non-zero minimal length are intertwined. On the
other hand, the Heisenberg uncertainty principle forms the core of
the EPR paradox. Subsequently, here, the implications of resorting
to the generalized uncertainty principle (or equally, the minimal
length) instead of the Heisenberg uncertainty principle on the
quantum non-locality are investigated through focusing on the
Franson experiment in which the uncertainty relation is the
backbone of understanding and explaining the results.
\end{abstract}
\maketitle
%
%%%%%%%%%%%%%%%%%%%%%%%%%%%%%%%%%%%%%%%%%%%%%%%%%%%%%%%%%%
\section{Introduction}

Certainly, quantum non-locality (QNL) is one of the most
intriguing subjects in physics rooted in the famous paper by
Einstein, Podolsky, and Rosen (EPR) \cite{EPR}. Basically, there
is a deep connection between QNL and the Heisenberg uncertainty
principle (HUP) \cite{EPR,Heisenberg,Franson,oppen}. Indeed, this
property has been obtained since people could reach quantum energy
levels. The quality of the validity of the Schr\"{o}dinger
equation (or equally, the quantum mechanics) decreases as the
energy of a system increases, and at high energy levels, quantum
mechanical interpretations should be replaced by those of the
quantum field theory. Therefore, it is significant challenge to
study the quality of non-locality in high energy physics to answer
the question of whether there is a change in QNL with increasing
energy or not? In this regard, the effects of special relativity
and curved spacetime on the behavior of QNL have extensively been
studied \cite{pres,tera,kim,fu,al,ver,ma,fris,moradpour2}.

%\cite{pres,20,21,tera,fu,al,ma,le,fun,smith,alsing,shi,ball,ver,refn,moradpour2,kim,fris}.

Despite the success of general relativity, the relation
between quantum mechanics and gravity is still mysterious
\cite{carlip}, and attempts to find a quantum gravity scenario
continue \cite{rovelli,seb}. One common feature of quantum gravity
(QG) scenarios is the existence of a non-zero minimum length, also
expectable in Newtonian gravity \cite{mead1}. Additionally,
the existence of such a non-zero minimum length naturally leads to
the generalized uncertainty principle (GUP) \cite{seb} meaning
that the quantum mechanical commutators of operators should change
if one wants to recover GUP. Therefore, this correspondence is a
great motivation to replace GUP with HUP, which stimulates us to
study the implications of this replacement on QNL.

In general, when the quantum features of gravity are considered,
canonical operators $x$ and $p$ are replaced with their
generalized counterparts $X$ and $P$, respectively, and up to the
first order of the GUP parameter $\beta$, we can write
$P_i=p_i(1+\beta f(p))$ in the position representation where
$X=x$. In order to get an insight into the implications of QG
(GUP) on the current physics, one may estimate the effects of QG
as the perturbations to the quantum mechanics, classical
mechanics, and quantum field theory
\cite{seb,prdgup,prdgup1,2,gupclass,fi1,rep,Ali:2011fa,fi2,fi5}.%\cite{seb,prdgup,prdgup1,2,Ali:2011fa,QIP,rep,Chung,gupclass,sarah,wave,fi1,fi2,fi3,fi4,fi5,fi6,fi7,fi8,fi9,fi10,fi11}.

As it has been argued, QNL is the result of HUP (the result of the
non-commutative of operators) \cite{EPR,Heisenberg,Franson,oppen}.
Dependency of the square of Bell's operators to the commutators is
a bright signal from it \cite{alsina,cerec}. This point has been
addressed in Ref.~\cite{QIP} where, considering the effects of GUP
on angular momentum algebra, it is shown that the Bell's operators
square of two partite systems changes. Based on this paper, the
Bell operator and thus its expectation value do not change as this
operator consists of operators with eigenvalues $\pm1$, and hence,
it remains unchanged. Therefore, while the Bell's operator does
not change, its square changes, an incompatibility. Indeed,
compared to the square of Bell's operator, the role of commutators
(or equally, HUP) in the original Bell's inequality is not obvious
and it is figured in the coincidence rate version appeared in the
Franson experiment \cite{PRL2009,Franson,pra2013}. Finally, the
fundamental question of whether GUP affects Bell's inequality and
QNL or not still needs to be studied.

Consequently, in order to answer the mentioned question, we focus
on the Franson experiment where the role of HUP is vital to
interpret the results. The paper is structured as follows. After
providing a general remarks on the Franson experiment in section
$(\textrm{II})$, the implications of the existence of a minimal
length on its outcome are addressed in Sec.~$(\textrm{III})$. A
summary has also been presented in the last section.
%%%%%%%%%%%%%%%%%%%%%%%%%%%%%%%%%%%%%%%%%%%%%%%%%%%%%%%%%%%%%%%%%%
\section{Franson Experiment}

There is a three levels quantum mechanical system in the Franson
setup, such that the highest energy level has energy $E_1$ and
relatively long lifetime $\tau_1$, the intermediate state of
energy $E_2$ and lifetime $\tau_2\ll\tau_1$, and the ground state
energy $E_3$ of very long lifetime $\tau_3$ ($\tau_1<\tau_3$)
\cite{Franson}. These states are labeled with states $\varphi_1$,
$\varphi_2$, and $\varphi_3$, respectively \cite{Franson}. In the
Franson experiment, uncertainty in the position of photons is
reflected in their transit time difference $\Delta T$ satisfying
$\tau_2\ll\Delta T\ll\tau_1$ (See Ref.~\cite{Franson} for more
info about the setup). Finally, the fields at the detectors $D_1$
and $D_2$ are written as \cite{Franson}

\begin{eqnarray}\label{11}
\varphi_{k}\left(\mathrm{x}_{1}, t\right)=\frac{1}{2}
\varphi_{k,0}\left(\mathrm{x}_{1}, t\right)+\frac{1}{2} e^{i
\phi_{1}} \varphi_{k,0}\left(\mathrm{x}_{1}, t-\Delta T\right),
\end{eqnarray}

\noindent and

\begin{eqnarray}\label{110}
\varphi_{k}\left(\mathrm{x}_{2}, t\right)=\frac{1}{2}
\varphi_{k,0}\left(\mathrm{x}_{2}, t\right)+\frac{1}{2} e^{i
\phi_{2}} \varphi_{k,0}\left(\mathrm{x}_{2}, t-\Delta T\right),
\end{eqnarray}

\noindent respectively. Here, $\phi_1$, and $\phi_2$ store the
information related to the phase shifts due to the half-silvered
mirrors $D_1$ and $D_2$, respectively, and $\Delta T$ is assumed
to be the same for both photons \cite{Franson}. In
Ref.~\cite{Franson}, $R_{c}$ (the coincidence rate between two
detectors) is evaluated by

\begin{eqnarray}
&&R_{c}=\\&&\eta_{1} \eta_{2}\left\langle
0\left|\varphi_{k}^{\dagger}\left(\mathbf{x}_{1}, t\right)
\varphi_{k}^{\dagger}\left(\mathbf{x}_{2}, t\right)
\varphi_{k}\left(\mathbf{x}_{2}, t\right)
\varphi_{k}\left(\mathbf{x}_{1}, t\right)\right|
0\right\rangle\nonumber,
\end{eqnarray}

\noindent where $\eta_{1},\eta_{2}$ denote the efficiency of the
corresponding detectors, and we can briefly write

\begin{eqnarray}
R_{c}&=&\frac{1}{16} \eta_{1} \eta_{2}\langle
0|A^{\dagger}A|0\rangle,
\end{eqnarray}

\noindent in which

\begin{eqnarray}
A&=&\varphi_{k,0}\left(\mathbf{x}_{1}, t\right) \varphi_{k,0}\left(\mathbf{x}_{2}, t\right)\nonumber\\
&+&\rm e^{i \phi_{1}} \rm e^{i \phi_{2}} \varphi_{k,0}\left(\mathbf{x}_{1}, t-\Delta T\right) \varphi_{k,0}\left(\mathbf{x}_{2}, t-\Delta T\right).
\end{eqnarray}

\noindent Whenever $\Delta T \ll \tau_{1}$, the amplitude of
detecting a pair of particles at time $t-\Delta T$ will be
approximately equal to the amplitude of detecting a pair of
particles at time $t$, and in fact, they have only a constant
phase difference causing \cite{Franson}

\begin{eqnarray}
&&\!\!\!\!\!\!\!\varphi_{k,0}\left(\mathbf{x}_{1}, t\right)
\varphi_{k,0}\left(\mathbf{x}_{2}, t\right)=\nonumber\\
&&\sum_{k_{1},k_{2}} c_{k_1} c_{k_2}\rm e^{i\left(\mathbf{k}_{1}
\cdot \mathbf{x}_{1}-\omega_{1} t\right)} \rm
e^{i\left(\mathbf{k}_{2} \cdot \mathbf{x}_{2}-\omega_{2}
t\right)},\\\nonumber &&\varphi_{k,0}\left(\mathbf{x}_{1}, t-\Delta
T\right)
\varphi_{k,0}\left(\mathbf{x}_{2}, t-\Delta T\right)=\\
&&\sum_{k_{1},k_{2}}  c_{k_1} c_{k_2} \rm
e^{i\left(\omega_{1}+\omega_{2}\right) \Delta T}\times \rm
e^{i\left(\mathbf{k}_{1} \cdot \mathbf{x}_{1}-\omega_{1} t\right)}
\rm e^{i\left(\mathbf{k}_{2} \cdot \mathbf{x}_{2}-\omega_{2}
t\right)}\nonumber,
\end{eqnarray}

\noindent where $c_{k_1},c_{k_2}$ are the expansion coefficients in
the Fourier transformation and can be determined by evaluation of
system. Energy conservation yields
$\omega_{1}+\omega_{2}=\left(E_{1}-E_{3}\right) / \hbar+\Delta
\omega$, where $E_{1}$ and $E_{3}$ are the unperturbed energies of
initial and final states, respectively. $\Delta\omega\sim
\frac{1}{\tau_{1}}+\frac{1}{\tau_{3}}$ (the uncertainty in
$\omega_{1}+\omega_{2}$), and its value is much less than the
individual uncertainty of $\omega_{i}$ since $\tau_2$ is
relatively short \cite{Franson}, and thus

\begin{eqnarray}
&&\varphi_{k,0}\left(\mathbf{x}_{1}, t-\Delta T\right) \varphi_{k,0}\left(\mathbf{x}_{2}, t-\Delta T\right) \\
&&=\rm e^{i\left(E_{1}-E_{3}\right) \Delta T / \hbar}
\varphi_{k,0}\left(\mathbf{x}_{1}, t\right)
\varphi_{k,0}\left(\mathbf{x}_{2}, t\right)\nonumber,
\end{eqnarray}

\noindent leading to

\begin{eqnarray}
R_c&=&\frac{1}{16} R_{0}\left[1+\rm e^{-i\left(\Delta E \Delta T/ \hbar+\phi_{1}+\phi_{2}\right)}\right]\times\nonumber\\
&&\left[1+\rm e^{i\left(\Delta E \Delta T / \hbar+\phi_{1}+\phi_{2}\right)}\right],
\end{eqnarray}

\noindent in which $R_0=\langle0|\sum_{k_{1},k_{2}} c^{\dagger}_{k_1}c^{\dagger}_{k_2}c_{k_1}c_{k_2}|0\rangle$ is the coincidence rate with the
half-silvered mirrors removed (shorter length) and $\Delta E=
E_1-E_3$. Finally, one finds \cite{Franson}

\begin{eqnarray}\label{1}
R_{c} &=&\frac{1}{4} R_{0} \cos ^{2}\left(\frac{\Delta E \Delta T / \hbar+\phi_{1}+\phi_{2}}{2}\right) \nonumber\\
&=&\frac{1}{4} R_{0} \cos ^{2}\left(\phi_{1}^{\prime}-\phi_{2}^{\prime}\right),
\end{eqnarray}

\noindent where

\begin{eqnarray}
&&\phi_{1}^{\prime}=\phi_{1} / 2,\\
&&\phi_{2}^{\prime}=-\left(\phi_{2}+\Delta E \Delta T /
\hbar\right) / 2\nonumber.
\end{eqnarray}

\section{Franson experiment in the presence of minimum length}

In this section, we intend to study the implications of the
quantum features of gravity on Eq.~(\ref{1}) and thus QNL using
the perturbation theory. Now, suppose that the quantum gravity
modifications to the two emitted photons in Franson experiment
have been considered. It means that the Hamiltonian of atoms and
thus their energy levels are also perturbed by the modifications
of the QG scenarios, and thus we have $\hat H_{GUP}=\hat H +\beta
\hat H_{p}$, where $\hat H_{p}$ refers to the perturbed
Hamiltonian in the GUP frame, and $E_{GUP}=E+\beta E_p$ for the
energy levels. $E_p$ can be determined using the perturbation
theory (up to the desired level).

Therefore, in the language of quantum field theory and due to the
existence of minimal length, the field operator of each photon is
modified as $\varphi^{GUP}_{k}(x)=\varphi_{k}(x)+\beta
\varphi^p_{k}(x)$, where the index $p$ denotes the correction
terms in the GUP framework \cite{seb,fi1}.
%\cite{seb,fi1,fi2,fi3,fi4,fi5,fi6,fi7,fi8,fi9,fi10,fi11}.
The time evolution of $\varphi^{GUP}_{k}(x)$ is obtained by

\begin{eqnarray}
\varphi^{GUP}_{k}(x, t)&=& \rm e^{i\hat H_{GUP}t/ \hbar} \varphi^{GUP}_{k}(x)\rm e^{-i\hat H_{GUP}t/ \hbar}\nonumber\\
&=&\varphi_{k}(x,t)-i\beta \Gamma_{k}(x,t)+\beta \varphi^{p}_{k}(x,t)\nonumber\\
&+&\mathcal{O}(\beta^{2}),
\end{eqnarray}

\noindent where $\Gamma_{k}(x,t)=[\varphi_{k}(x,t),\hat H_{p}]t/
\hbar$.

For the counterparts of Eqs.~(\ref{11}) and~(\ref{110}), similar
to the above argument, and by following the Franson approach, the
field corresponding to the $i$th photon, at the detector $D_i$ can
be written as

\begin{eqnarray}\label{0index}
\varphi^{GUP}_{k}\left(\mathrm{x}_{i}, t\right)&=&\frac{1}{2} \varphi^{GUP}_{k,0}\left(\mathrm{x}_{i}, t\right)\nonumber\\
&+&\frac{1}{2} \rm e^{i \phi_{1}}
\varphi^{GUP}_{k,0}\left(\mathrm{x}_{i}, t-\Delta T\right).
\end{eqnarray}

\noindent Finally, the corresponding coincidence rate $R^{QG}_{c}$
is achieved by

\begin{eqnarray}
R^{QG}_{c}&=&\eta'_{1}\eta'_{2}\langle
0|\varphi_{k}^{GUP,\dagger}\left(\mathbf{x}_{1}, t\right)
\varphi_{k}^{GUP,\dagger}\left(\mathbf{x}_{2}, t\right)\nonumber\\
&\times&\varphi^{GUP}_{k}\left(\mathbf{x}_{2}, t\right) \varphi^{GUP}_{k}\left(\mathbf{x}_{1},
t\right)|0\rangle,
\end{eqnarray}

\noindent summarized into

\begin{eqnarray}
R^{QG}_{c}&=& \eta'_{1} \eta'_{2}\langle 0|B^{\dagger}B|0\rangle,
\end{eqnarray}

\noindent in which

\begin{eqnarray}
\!\!\!\!\!\!&B&=\varphi_{k}(x_2,t)\varphi_{k}(x_1,t)\\
\!\!\!\!\!\!&+&\frac{1}{2} \beta \bigg[\varphi^{p}_{k,0}(x_2,t)\varphi_{k}(x_1,t)+\rm e^{i\phi_{2}}\varphi^{p}_{k,0}(x_2,t-\Delta T)\varphi_{k,0}(x_1,t)\nonumber\\
\!\!\!\!\!\!&-&i\Gamma_{k,0}(x_2,t)\varphi_{k,0}(x_1,t)-i\rm e^{i\phi_{2}}\Gamma_{k,0}(x_2,t-\Delta T)\varphi_{k,0}(x_1,t)\nonumber\\
\!\!\!\!\!\!&+&\varphi_{k}(x_2,t)\varphi^{p}_{k,0}(x_1,t)+\rm e^{i\phi_{1}}\varphi_{k}(x_2,t)\varphi^{p}_{k,0}(x_1,t-\Delta T)\nonumber\\
\!\!\!\!\!\!&-&i\varphi_{k}(x_2,t)\Gamma_{k,0}(x_1,t)-i\rm
e^{i\phi_{1}}\varphi_{k,0}(x_2,t)\Gamma_{k,0}(x_1,t-\Delta
T)\bigg].\nonumber
\end{eqnarray}

\noindent Using the Fourier expansion, one can write

\begin{eqnarray}
&&\varphi^{p}_{k,0}\left(\mathbf{x}_{1}, t\right)
\varphi^{p}_{k,0}\left(\mathbf{x}_{2}, t\right)=
\nonumber\\&&\sum_{k_{1},k_{2}} c'_{k_1}c'_{k_2}\rm
e^{i\left(\mathbf{k}_{1} \cdot \mathbf{x}_{1}-\omega_{1} t\right)}
\rm e^{i\left(\mathbf{k}_{2} \cdot \mathbf{x}_{2}-\omega_{2}
t\right)},
\end{eqnarray}

\noindent and

\begin{eqnarray}
&&\Gamma_{k,0}\left(\mathbf{x}_{1}, t\right)
\Gamma_{k,0}\left(\mathbf{x}_{2}, t\right)=
    \nonumber\\&&\sum_{k_{1},k_{2}} c''_{k_1}c''_{k_2}\rm
    e^{i\left(\mathbf{k}_{1} \cdot \mathbf{x}_{1}-\omega_{1} t\right)}
    \rm e^{i\left(\mathbf{k}_{2} \cdot \mathbf{x}_{2}-\omega_{2}
        t\right)},
\end{eqnarray}

\noindent leading to

\begin{eqnarray}
&&\varphi^{p}_{k,0}\left(\mathbf{x}_{1}, t-\Delta T\right)
\varphi^{p}_{k,0}\left(\mathbf{x}_{2}, t-\Delta
T\right)=\\&&\sum_{k_{1},k_{2}}c'_{k_1}c'_{k_2} \rm
e^{i\left(\omega_{1}+\omega_{2}\right) \Delta T}\rm
e^{i\left(\mathbf{k}_{1} \cdot \mathbf{x}_{1}-\omega_{1} t\right)}
\rm e^{i\left(\mathbf{k}_{2} \cdot
\mathbf{x}_{2}-\omega_{2}t\right)}.\nonumber
\end{eqnarray}

\noindent Here, $c'_{k_1},c'_{k_2}$ and $c''_{k_1},c''_{k_2}$ are the
corresponding coefficients in the Fourier expansion.

In this manner, the corresponding energy conservation leads to

\begin{eqnarray}
\omega_{1}+\omega_{2}&=&\Delta E/\hbar+\beta \Delta E_{p}/\hbar,
\end{eqnarray}

\noindent where $\Delta E=E_3-E_1$, and $\Delta
E_p=E_{3,p}-E_{1,p}$ which yields

\begin{eqnarray}
R^{QG}_c&=&\frac{1}{16} \bigg(R_{0}+2\beta(R'_1+R'_2)\bigg)\nonumber\\
&\times&\left[1+\rm e^{-i\left(\Delta E \Delta T/ \hbar+\beta\Delta E_p \Delta T/ \hbar+\phi_{1}+\phi_{2}\right)}\right]\nonumber\\
&\times&\left[1+\rm e^{i\left(\Delta E \Delta T /
\hbar+\beta\Delta E_p \Delta T/
\hbar+\phi_{1}+\phi_{2}\right)}\right],
\end{eqnarray}

\noindent and thus

\begin{eqnarray}
R^{QG}_{c} &=&\frac{1}{4} R^{GUP}_{0} \cos ^{2}\left(\frac{\Delta E \Delta T / \hbar+\beta\Delta E_p \Delta T/ \hbar  +\phi_{1}+\phi_{2}}{2}\right) \nonumber\\
&=&\frac{1}{4} R^{GUP}_{0} \cos ^{2}\left(\Phi_{1}^{\prime}-\Phi_{2}^{\prime}\right),
\end{eqnarray}
\\
\noindent where $R^{GUP}_0=R_{0}+2\beta(R'_1+R'_2)$ is the
coincidence rate of the shorter length near to the Planck scale,
and

\begin{eqnarray}
R'_1&=&\langle0|\sum_{k_{1},k_{2}} c^{\dagger}_{k_1}c^{\dagger}_{k_2}c_{k_1}c'_{k_2}|0\rangle,\nonumber\\
R'_2&=&\langle0|\sum_{k_{1},k_{2}} c^{\dagger}_{k_1}c^{\dagger}_{k_2}c'_{k_1}c_{k_2}|0\rangle,\nonumber\\
\Phi_{1}^{\prime}&=&\phi_{1} / 2,\\
\Phi_{2}^{\prime}&=&-\left(\phi_{2}+\Delta E \Delta T / \hbar+\beta\Delta E_p \Delta T/ \hbar\right) / 2.\nonumber
\end{eqnarray}

\noindent It is obvious that, at the limit of
$\beta\longrightarrow0$, the desired results obtained in quantum
mechanics are recovered. Therefore, the coincidence rate stores
the effects of QG, and indeed, the existence of minimum length
affects the spectrum of the coincidence rate and thus the spectrum
of the expectation value of Bell's operator.

%%%%%%%%%%%%%%%%%%%%%%%%%%%%%%%%%%%%%%%%%%%%%%%%%%%%%%%%%%%%%%%%%%%%
\section{Summary}

It seems that the existence of a non-zero minimal length is
unavoidable leading to GUP (or equally, modified commutators
algebra) \cite{seb,mead1}. Consequently, motivated by the deep
connection between HUP and the EPR paradox leading to the
emergence of QNL, and also Ref.~\cite{QIP}, showing that the
square of Bell's operators formed by angular momentum operators,
changes when HUP is replaced by GUP, we tried to clarify the
relation between QNL and GUP. In order to achieve this goal, we
resorted to the Franson experiment in which the uncertainty
principle plays a crucial role in describing the results, and
obtained that GUP affects the coincidence rate spectrum.

%%%%%%%%%%%%%%%%%%%%%%%%%%%%%%%%%%%%%%%%%%

\end{document}